\documentclass[final,5p,times,twocolumn]{elsarticle}

\usepackage{graphicx}

\usepackage{amssymb}
\usepackage{amsmath}
\usepackage{color}

\biboptions{sort&compress}
\journal{Applied Surface Science}

\def\CO2{CeO$_2$}
\def\C2O3{Ce$_2$O$_3$}

\def\LP{lattice parameter}
\def\LPs{lattice parameters}
\def\BM{bulk modulus}
\def\BMi{bulk moduli}

\newcommand{\corr}[1]{\color{black}{#1}\color{black}}
\newcommand{\corrfig}[1]{\color{black}{#1}\color{black}}

\begin{document}
\begin{frontmatter}

\title{Tuning of \CO2 buffer layers for coated superconductors through doping}
\author[SCRiPTS,QChem]{Danny E.\ P.\ Vanpoucke}
\ead{Danny.Vanpoucke@Ugent.be}
\author[CMM,MSE]{Stefaan Cottenier}
\author[CMM]{Veronique Van Speybroeck}
\author[QChem]{Patrick Bultinck}
\author[SCRiPTS]{Isabel Van Driessche}

\address[SCRiPTS]{SCRiPTS group, Department of Inorganic and Physical Chemistry, Ghent University, Krijgslaan $281$ - S$3$, BE-$9000$ Gent, Belgium}
\address[QChem]{Ghent Quantum Chemistry group, Department of Inorganic and Physical Chemistry, Ghent University, Krijgslaan $281$ - S$3$, BE-$9000$ Gent, Belgium}
\address[CMM]{Center for Molecular Modeling, Ghent University, Technologiepark $903$, BE-$9052$ Zwijnaarde, Belgium}
\address[MSE]{Department of Materials Science and Engineering, Ghent University, Technologiepark $903$, BE-$9052$ Zwijnaarde, Belgium }

\begin{abstract}
The appearance of microcracks in CeO$_2$ buffer layers, as used in buffer layer architectures for coated superconductors, indicates the presence of stress between this buffer layer and the substrate. This stress can originate from the differences in thermal expansion or differences in lattice parameters between the CeO$_2$ buffer layer and the substrate. In this article, we study, by means of \textit{ab initio} density functional theory calculations, the influence of group IV doping elements on the lattice parameter and bulk modulus of CeO$_2$. Vegard's law behavior is found for the lattice parameter in systems without oxygen vacancies, and the Shannon crystal radii for the doping elements are retrieved from the lattice expansions. We show that the lattice parameter of the doped CeO$_2$ can be matched to that of the La$_2$Zr$_2$O$_7$ coated NiW substrate substrate for dopant concentrations of about $5$\%, and that bulk modulus matching is either not possible or would require extreme doping concentrations.
\end{abstract}


\end{frontmatter}

\section{Introduction}\label{ASS:Introduction}
\indent Cerium-oxide-based materials have attracted increasing interest over the last two decades. This is mainly due to their remarkable properties with regard to oxidation-reduction catalysis. They are used in a number of industrial applications: three-way catalysts \cite{DeganelloF:SSI2002}, oxygen sensors, solid-oxide fuel cells \cite{TullerHL:JES1975,MikiTakeshi:JPC1990}, and many more \cite{KundakovicLj:JCat1998,SheYusheng:IJHE2009,ManzoliMaela:CT2008}. More recently, cerium oxide (CeO$_2$) has been used as thin film buffer layer in YBa$_2$Cu$_3$O$_{7-\delta}$ (YBCO) coated superconductors (CSC) \cite{ParanthamanM:1997PhysC,OhSanghyun:PhysC1998,PennemanG:2004EuroCeram,TakahashiY:PhysC2004,KnothKerstin:PhysC2005,VandeVeldeNigel:EurJInorChem2010}. In a YBCO-CSC architecture, a YBCO thin film is grown on a metallic substrate. To prevent the metal atoms of diffusing into the YBCO, one or more buffer layers are required. In addition, these buffer layers also prevent the oxidation of the metallic substrate during YBCO deposition. Due to its structural compatibility with YBCO, CeO$_2$ is preferred as the top layer in a multilayer architecture. However, the layer thickness of the CeO$_2$ buffer layer
is limited by the formation of cracks during deposition \cite{ParanthamanM:1997PhysC,OhSanghyun:PhysC1998}. This phenomenon has been linked to internal stress due to lattice mismatch or different thermal expansion coefficients of the substrate and the CeO$_2$ buffer layer \cite{OhSanghyun:PhysC1998,VandeVeldeNigel:EurJInorChem2010}. A simple way to reduce the mismatch and stress is through doping \cite{TakahashiY:PhysC2004,KnothKerstin:PhysC2005,VandeVeldeNigel:EurJInorChem2010}.\\
\indent In this paper, we study the influence of doping on this mismatch and stress using \textit{ab initio} atomistic calculations. The energetics and electronic properties of doped systems and the influence of oxygen vacancies is beyond the scope of this work and will be discussed elsewhere.\\
\indent The stress due to volumetric changes, which are present during the heating and cooling cycles of the production process, are investigated through the bulk moduli (BM) of the doped systems. Because the CeO$_2$ buffer layer is often grown on a La$_2$Zr$_2$O$_7$ (LZO) buffer layer in multilayer architectures, a match with the LZO bulk modulus will reduce the inter-layer stress. The change in the CeO$_2$ lattice parameter is derived directly from the calculated atomic structure, and a match with the LZO \LP\ is searched for to reduce inter-layer stress due to lattice mismatch.\\
\indent Because Ce is tetravalent in CeO$_2$, group IV elements are an obvious choice as doping elements. They have the additional advantage that no extra oxygen vacancies need to be introduced for charge compensation. This allows us to retain a clear picture of the direct effects on the \LP\ and \BM\  due to the doping elements themselves.\\
\indent We also look at the effect of aliovalent dopants (Cu, Zn, and La), assuming oxidized systems without the presence of charge compensating vacancies. This corresponds with experimental systems under (strongly) oxidizing atmosphere. Comparison to results of the group IV elements shows that the valency has relatively little influence.\\
\indent For such homogeneous systems without oxygen vacancies, it is possible to study the crystal structure from a purely analytic perspective, and we derive Vegard's empirical law analytically for these systems.
\section{Theoretical method}\label{ASS:TheoreticalMethod}
\begin{figure}[!tb]
\begin{center}
  \includegraphics[width=7cm,keepaspectratio=true]{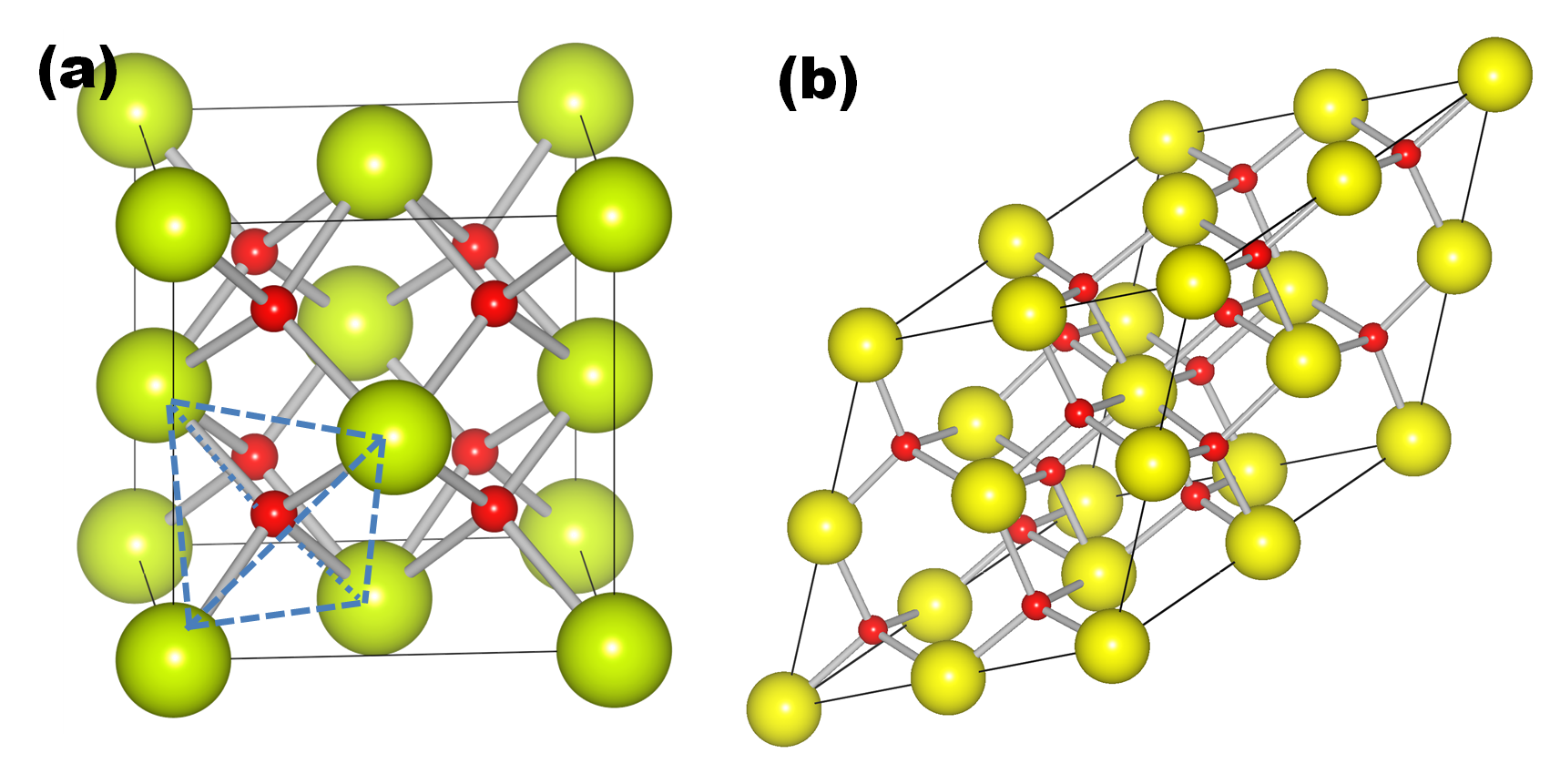}\\
  \caption{Ball-and-stick model presenting the CeO$_2$ cubic fluorite structure supercells (a) c111 and (b) p222 (\textit{cf.} text). The big yellow and small red spheres indicate the positions of the Ce and O atoms, respectively. The tetrahedral surrounding of a single O atom is indicated by the tetrahedron for the c111 supercell.}\label{fig:c111geom}
\end{center}
\end{figure}

\indent We perform \textit{ab initio} density functional theory (DFT) calculations within the projector augmented-wave method as implemented in the Vienna \textit{ab initio} Package (\textsc{VASP}) program \cite{Blochl:prb94,Kresse:prb99}. The local density approximation (LDA) functional as parameterized by Ceperley and Alder and the generalised gradient approximation (GGA) functional as constructed by Perdew, Burke and Ernzerhof (PBE) are used as exchange-correlation functionals \cite{CA:prl1980,PBE_1996prl,Kresse:prb93,Kresse:prb96}. Since our focus goes mainly to the mechanical and structural properties of the system this should be sufficient, and no additional Coulomb correction is required. The plane wave kinetic energy cutoff is set to $500$ eV.\\
\indent Symmetric supercells, containing a single dopant per supercell are used to simulate homogeneous distributions of the dopants. The supercells used are the fluorite cubic $1\!\times\!1\times\!1$ cell with $12$ atoms (c$111$), shown in Fig.~\ref{fig:c111geom}a, the primitive $2\!\times\!2\times\!2$ cell with $24$ atoms (p$222$), shown in Fig.~\ref{fig:c111geom}b, the primitive $3\!\times\!3\times\!3$ cell with $81$ atoms (p$333$)\corr{, not shown,}\ and the cubic $2\!\times 2\!\times\!2$ cell with $96$ atoms (c$222$)\corr{, not shown}. Replacing a single Ce atom results in dopant concentrations of $25.0, 12.5, 3.7,$ and $3.1$ \%, respectively. Monkhorst-Pack special $k$-point grids are used to sample the Brillouin zone \cite{Monkhorst:prb76}. For the two smaller cells we use an $8\!\times \!8\times\!8$ $k$-point grid while for the two large supercells a $4\!\times\!4\times\!4$ $k$-point grid is used. To optimize the structures, a conjugate gradient method is used. During relaxation both atom positions and cell geometry are allowed to change simultaneously. The convergence criterion is set to the difference in energy between subsequent steps becoming smaller than $1.0\times10^{-6}$ eV.\\
\indent For each dopant the \BM\  is calculated by fitting $E(V)$ data from fixed volume calculations to the third order isothermal Birch-Murnaghan equation of state \cite{MurnaghanFD:PNAS1944,BirchF:PhysRev1947}. To reduce the computational cost, dopant concentrations of $25$\% are used.\\
\indent Ball-and-stick images of the crystal structures are generated using the VESTA visualization tool \cite{VESTA:JApplCryst2008}.

\section{Results and Discussion}\label{ASS:ResultDiscuss}
\subsection{Analytic derivation of Vegard's Law}
\indent CeO$_2$ is known to have a cubic fluorite structure (space group $Fm\overline{3}m$). Figure~\ref{fig:c111geom}a shows the c111 supercell. Since every O atom is tetrahedrally surrounded by Ce atoms, the following relation can be derived from the Ce--O bond length:
\begin{equation}\label{eq:Rce_def}
R_{O}+R_{Ce}=a_{CeO_2}\frac{\sqrt{3}}{4},
\end{equation}
with $R_{O}$ and $R_{Ce}$ the atomic crystal radii of O and Ce respectively, and $a_{CeO_2}$ the CeO$_2$ \LP. For a cubic system, this allows us to calculate the atomic crystal radius $R_{M}$ of a Ce substituent $M$ via:
\begin{equation}\label{eq:Rm_def}
R_{M} = \Bigg(\frac{\sqrt{3}}{4}a_{M_xCe_{1-x}O_{2}}-R_{O}-(1-n_x)R_{Ce}\Bigg)/n_x
\end{equation}
with $a_{M_xCe_{1-x}O_2}$ the \LP\ of the doped system and $n_x$ the dopant concentration. The \LP\ of the doped system can now be found by combining Eqns.~(\ref{eq:Rce_def}) and (\ref{eq:Rm_def}):
\begin{equation}\label{eq:vegards_in_CeO2M}
a_{M_xCe_{1-x}O_2} = a_{CeO_2} + \Bigg(\frac{4}{\sqrt{3}}(R_{O}+R_{M})-a_{CeO_2}\Bigg)n_x.
\end{equation}
This results in a clear linear relation between the lattice expansion/contraction of CeO$_2$ and the dopant concentration which is known as Vegard's empirical law \cite{VegardsLaw_DentonAR:PhysRevA1991}.
These analytical results show Vegard's law behavior should be expected when the dopants are homogeneously distributed in CeO$_2$ in case of tetravalent dopants, but also for non-tetravalent dopants this behavior should be expected under oxidizing conditions.\\
\indent Although Eqns.~(\ref{eq:Rm_def}) and (\ref{eq:vegards_in_CeO2M}) can be derived from one another, the latter is more interesting from the experimental point of view, since \corr{\LPs}\ and concentrations are readily available while atomic radii are not. Values for the two radii $R_O$ and $R_M$ can be taken from tabulated values for atomic radii. This, however, can be problematic since there are several different definitions for `atomic radius' available, giving values which can easily differ 20\%.\footnote{\textit{E.g.} for oxygen one finds the calculated atomic radius to be $0.48$ \AA\ while the empirical atomic radius is given to be $0.60$ \AA\  \cite{ClementiE:1963JChemPhys_webelementsAtomicRadiiCalculated, SlaterJC:JChemPhys1964_webelementsAtomRadiiEmpirical}.} In light of this problem, Eqn.~(\ref{eq:Rm_def}) becomes interesting. It could tell us which definition of atomic radius to use for the doped CeO$_2$ systems. This would allow one to predict the Vegard's law behavior for any doped CeO$_2$ system prior to its synthesis.

\begin{table*}[!tb]
\caption{Dopant radii calculated using Eqn.~(\ref{eq:Rm_def}), averaged over the four dopant concentrations (avg), and standard deviation (stdev) of this value. This is done for both LDA and PBE calculated geometries. The Shannon crystal radii for the 8-coordinated tetravalent atoms $R_{sh}^{8}$, taken from Ref.~\cite{Shannon:table}, are shown in comparison. $a_0$ and $b$ are the intercept and slope of the Vegard's law linear fit to the calculated geometries for doped CeO$_2$ systems. The \BMi\ are calculated for dopant concentrations of $25$\%. The \LP\ and \BM\ (BM) calculated for pure CeO$_2$ and La$_2$Zr$_2$O$_7$(LZO) are given as reference.}\label{table:MetalX_Rm_BM_Vegards}
\begin{center}
\begin{tabular}{l|ccccc|crcrcc}
 & \multicolumn{4}{c}{R$_M$ (\AA)} & $R_{sh}^{8}$ & \multicolumn{4}{c}{Vegard's Law} & \multicolumn{2}{c}{BM (Mbar)} \\
 & \multicolumn{2}{c}{LDA} & \multicolumn{2}{c}{PBE} & & \multicolumn{2}{c}{LDA} & \multicolumn{2}{c}{PBE} & LDA & PBE \\
 & avg & stdev & avg & stdev & (\AA)& $a_0$(\AA) & \multicolumn{1}{c}{$b$} & $a_0$(\AA) & \multicolumn{1}{c}{$b$} &  &  \\
\hline\\[-2mm]
CeO$_2$ & $1.0819$$^a$ & $0.0001$ & $1.1257$$^a$ & $0.0004$ & $1.11$ & \multicolumn{2}{c}{$5.3623^{b}$} & \multicolumn{2}{c}{$5.4629^{b}$} & $2.017$ & $1.715$ \\
LZO & $$ & $$ & $$ & $$ & $$ & \multicolumn{2}{c}{$10.6923^{c}$} & \multicolumn{2}{c}{$10.8906^{c}$} & $1.774$ & $1.542$ \\
\hline\\[-2mm]
C   & $0.9243$ & $0.0137$ & $1.0096$ & $0.0130$ & $-$ & $5.3656$ & $-0.4161$ & $5.4657$ & $-0.3077$ & $1.528$ & $1.235$ \\
Si  & $0.7951$ & $0.0052$ & $0.8321$ & $0.0044$ & $-$ & $5.3626$ & $-0.6688$ & $5.4643$ & $-0.6936$ & $2.057$ & $1.738$ \\
Ge  & $0.8786$ & $0.0031$ & $0.9270$ & $0.0034$ & $-$ & $5.3618$ & $-0.4640$ & $5.4631$ & $-0.4578$ & $1.909$ & $1.573$ \\
Sn  & $0.9764$ & $0.0020$ & $1.0199$ & $0.0049$ & $0.95$ & $5.3618$ & $-0.2383$ & $5.4629$ & $-0.2400$ & $2.004$ & $1.692$ \\
Pb  & $1.0686$ & $0.0034$ & $1.1293$ & $0.0042$ & $1.08$ & $5.3612$ & $-0.0174$ & $5.4633$ & $ 0.0050$ & $1.845$ & $1.516$ \\
\hline\\[-2mm]
Ti  & $0.8421$ & $0.0043$ & $0.8862$ & $0.0050$ & $0.88$ & $5.3629$ & $-0.5640$ & $5.4644$ & $-0.5706$ & $2.145$ & $1.825$ \\
Zr  & $0.9548$ & $0.0019$ & $0.9791$ & $0.0052$ & $0.98$ & $5.3622$ & $-0.2938$ & $5.4634$ & $-0.3409$ & $2.153$ & $1.878$ \\
Hf  & $0.9205$ & $0.0023$ & $0.9612$ & $0.0056$ & $0.97$ & $5.3622$ & $-0.3733$ & $5.4635$ & $-0.3849$ & $2.194$ & $1.881$ \\
\hline\\[-2mm]
Cu  & $0.9131$ & $0.0024$ & $0.9907$ & $0.0064$ & $0.91$$^e$/$-$$^h$ & $5.3624$ & $-0.3947$ & $5.4627$ & $-0.3065$ & $1.704$ & $1.374$ \\
  & $$ & $$ & $$ & $$ & $$ & $$ & $$ & $$ & $$ & $1.867$$^d$ & $1.553$$^d$ \\
Zn  & $0.9518$ & $0.0046$ & $1.0277$ & $0.0077$ & $0.88$$^f$/$1.04$$^i$ & $5.3632$ & $-0.3173$ & $5.4640$ & $-0.2389$ & $1.712$ & $1.410$ \\
La  & $1.1858$ & $0.0011$ & $1.2421$ & $0.0043$ & $1.17$$^g$/$1.30$$^j$ & $5.3623$ & $0.2374$ & $5.4637$ & $0.2597$ & $1.835$ & $1.556$ \\
\end{tabular}
\end{center}
\begin{flushleft}
$^a$ The Ce radius is calculated using Eqn.~(\ref{eq:Rce_def}), where the $4$-coordinated Shannon crystal radius for oxygen is taken as $1.24$ \AA\ \cite{Shannon:table}.\\
$^b$, $^c$ The actual \LP\ as calculated from the pure, relaxed geometries of CeO$_2$ and LZO. Note that the LZO \LP\ is double the CeO$_2$ \LP.\\
$^d$ The \BM\ for Cu dopant concentration of $12.5$\%.\\
$^e$, $^f$, $^g$ Shannon crystal radii for 6-coordinated Cu$^{1+}$, Zn$^{2+}$, and La$^{3+}$.\\
$^h$, $^i$, $^j$ Shannon crystal radii for 8-coordinated Cu$^{1+}$, Zn$^{2+}$, and La$^{3+}$.
\end{flushleft}
\end{table*}

\begin{figure}[!tb]
\begin{center}
  \includegraphics[width=8.5cm,keepaspectratio=true]{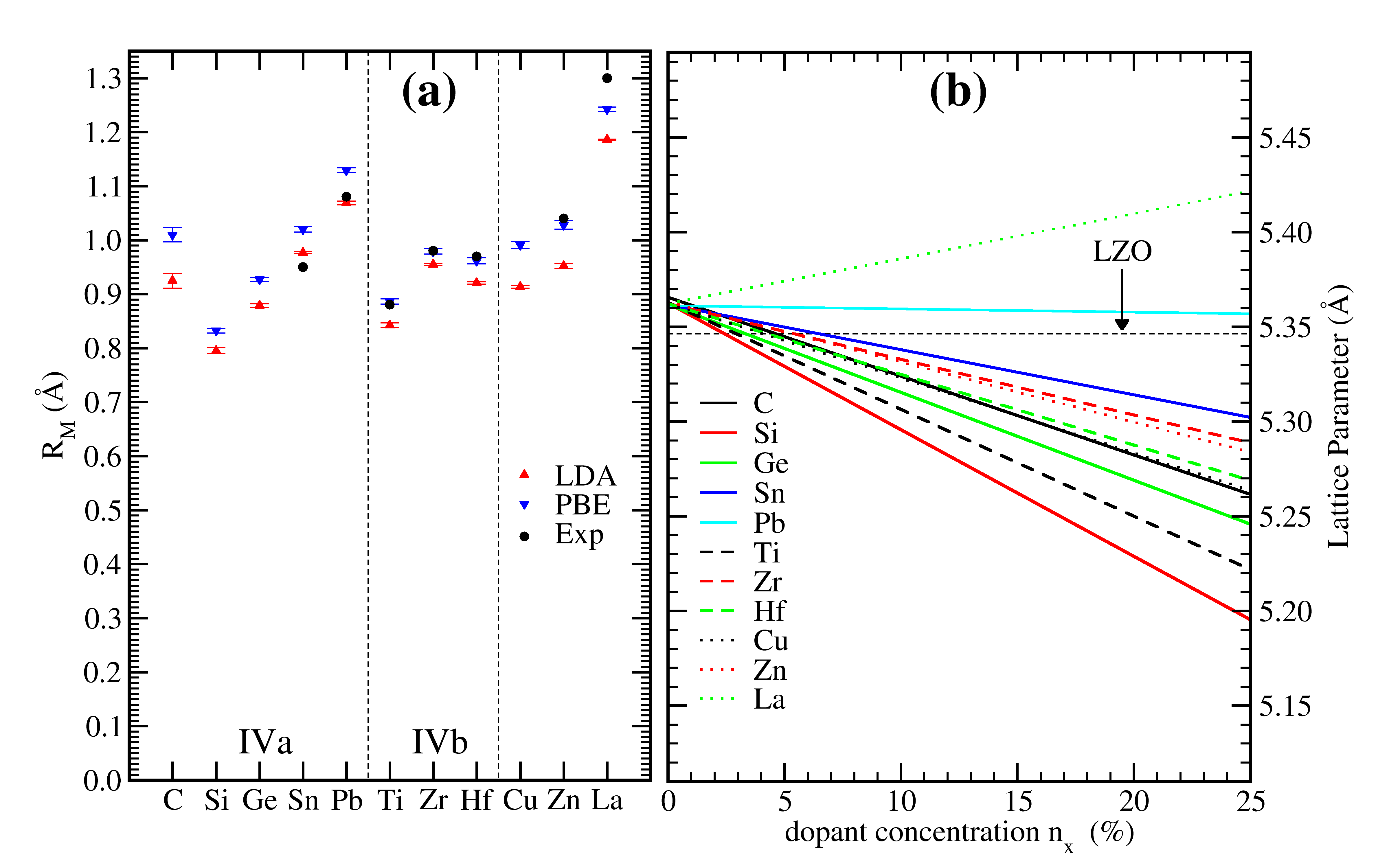}\\
  \caption{\corrfig{(a) The dopant radii calculated using Eqn.~\ref{eq:Rm_def} for both the LDA and PBE results. The standard deviation is shown as error bars, and, where available, the experimental Shannon crystal radii $R_{sh}^{8}$ (\textit{cf.}~Table~\ref{table:MetalX_Rm_BM_Vegards}) are shown as reference. (b) The \LP\ as function of the dopant concentration, calculated using Eqn.~\ref{eq:vegards_in_CeO2M}. Only the LDA results are shown. The LZO half lattice parameter is shown as reference.}}\label{fig:LPvsNx_and_Rm}
\end{center}
\end{figure}
\subsection{Group IV elements}
\indent Using Eqn.~(\ref{eq:Rm_def}) we have calculated the atomic crystal radius for each of the four concentrations for each group IV element. The average values and their standard deviations are shown in Table~\ref{table:MetalX_Rm_BM_Vegards} \corrfig{and Fig.~\ref{fig:LPvsNx_and_Rm}a}. In these calculations we have used the Shannon crystal radius for four-coordinate O$^{2-}$; $R_O=1.24$ \AA. The radius of eight-coordinate Ce$^{4+}$ is calculated from the non-doped CeO$_2$ system using Eqn.~(\ref{eq:Rce_def}) and shows very good agreement with the Shannon crystal radius for eight-coordinate Ce$^{4+}$\cite{Shannon:ACSA1976,Shannon:table}. For the group IV elements, the LDA and PBE calculations show the same relative trends, with the PBE values always slightly larger than the LDA ones, as one would expect. Due to the underbinding nature of GGA functionals such as PBE and the overbinding nature of LDA, we expect the LDA and PBE values to be a slight under- and overestimation of the actual crystal radius, respectively. Comparison of the calculated dopant radii to the Shannon crystal radii for eight-coordinated tetravalent atoms $R_{sh}^8$ shows a very good correlation. This indicates that the \emph{Shannon crystal radius} is an excellent parametrization to predict the lattice expansion of doped CeO$_2$.\\
\indent Based on \corr{\LPs}\ obtained from the \emph{ab initio} calculations for the different dopant concentrations, a fitting of Vegard's law is done for each of the group IV elements. Table~\ref{table:MetalX_Rm_BM_Vegards} shows the intercept $a_0$ and the slope $b$ of this linear fitting. \corrfig{For the LDA results, the fitted curves of the \LP\ as function of the dopant concentration $n_x$ are shown in Fig.~\ref{fig:LPvsNx_and_Rm}b.}\ With the exception of Pb, all systems show a nearly perfect fit, with \corr{correlation coefficient}\ ($R^2$) values better than $0.99$ for both LDA and PBE calculations. The poor fit of Pb is related to the negligible expansion of the lattice. As a result, small deviations can transform an expansion into a contraction going from one concentration to the next. A zero expansion can thus be assumed for Pb doping. The high quality of the other linear fits, with intercepts that are within $0.01$ \AA\ of the CeO$_2$ \LP, shows that the analytically obtained Vegard's law is a good model for the lattice expansion in these doped systems.\\
\indent Both LDA and PBE results show the same qualitative behavior, and the
optimum substitution concentrations, for \LP\ matching with LZO, are $\sim5$\% for all group IV elements, with, due to its negligible contraction, the exception of Pb substitution.\\
\indent Table~\ref{table:MetalX_Rm_BM_Vegards} also shows the \BM, obtained for substituent concentrations of $25$\%. If we assume the \BM\  to behave linearly with regard to the dopant concentration (\textit{cf.}~next section), then it is possible to estimate the optimum dopant concentration which would result in a perfect matching of the \BM\  of the doped CeO$_2$ and LZO. For the elements of group IVb, table~\ref{table:MetalX_Rm_BM_Vegards} shows a \BM\  which is larger than that of CeO$_2$. This makes \BM matching with LZO, which has a smaller \BM, impossible. In addition, the \BM\ seems to increase very slightly with increasing atomic number. The group IVa elements show a more complex behavior. With the exception of Si, all group IVa elements lower the \BM\ of CeO$_2$. However, the effect is generally too small to allow for a \BM\ matching at reasonable dopant concentrations. In case of the IVa elements, the \BM\ is lowered with the introduction of every newly filled shell ($d$ for Ge and Sn, and $f$ for Pb), while it increases with increasing atomic radii. Combining the results of the groups IVa and IVb shows that not only the valency electrons but also electrons in filled shells near the Fermi level play a crucial role for the \BM.
\subsection{Aliovalent dopants without vacancies}
\indent We have also investigated the effect of doping with aliovalent elements, but this without the introduction of charge compensating vacancies. This makes comparison between aliovalent and group IV elements more straightforward. The three aliovalent elements we have investigated are: Cu, Zn, and La. (The effect of the addition of charge compensating vacancies is a topic on its own, and will therefore be discussed elsewhere.)\\
\indent If we look at the calculated atomic radii in Table~\ref{table:MetalX_Rm_BM_Vegards}, we see that unlike with the group IV elements, the values are smaller than the Shannon crystal radii for 8-coordinated atoms, but larger than the values for 6-coordinated atoms.\footnote{Extrapolating the tabulated Shannon crystal radius values for Cu$^{1+}$ to the coordination number $8$ would give a value in the range of $1.05$--$1.11$ \AA.} This would indicate that the lower valence results in a lower coordination, regardless of the geometric and chemical surrounding. This in turn should give rise to charge redistribution near the defect.\\
\indent Just as for the group IV elements, the obtained \corr{\LPs}\ for different concentrations can be fitted nicely against a linear Vegard law, with $R^2$ values better than $0.99$. This means that under oxidizing atmosphere, \textit{i.e.} with little or no oxygen vacancies in the system, also aliovalent dopants should show a linear concentration dependence of the \LP. As for the group IV elements \LP\ matching with LZO is found to be $\sim5$ \% for Cu and Zn. For La doping, which shows a lattice expansion, no \LP\ matching with LZO is possible.\\
\indent In case of Cu, we calculated the \BM\ for dopant concentrations of $25.0$ and $12.5$\%, shown in Table~\ref{table:MetalX_Rm_BM_Vegards}. Taking the CeO$_2$ \BM\ as the case of $0.0$\% doping, a nearly linear trend is observed. Although all three aliovalent elements show a decrease of the \BM\  which looks slightly better than is the case for the group IV elements, very high dopant concentrations would still be needed to have \BM\ matching with LZO.\\
\indent Combined with the results for the group IV elements, it is clear that the \LP\ is only influenced by the (Shannon) crystal radius of the doping elements, while the \BM\ is also strongly influenced by the electronic structure (compare Zn/Zr and Hf/Cu).
\section{Conclusion}\label{ASS:Conclusion}
\indent The influence of group IV and aliovalent dopants on the \LP\ and \BM\ of CeO$_2$ are investigated using DFT calculations. A Vegard's law relation is analytically derived for doped CeO$_2$ without oxygen vacancies. \corr{\LPs}\  obtained from DFT calculations for different dopant concentrations show that for both group IV and aliovalently doped systems without oxygen vacancies the lattice expansion is described by a linear relation. This Vegard's law can be predicted from the Shannon crystal radius of the dopant element. Optimum doping concentrations for \LP\ matching with LZO is about $5$ \% for the different dopants studied, the exceptions being Pb and La. The former shows no appreciable expansion of contraction of the CeO$_2$ \LP, while the latter results in a lattice expansion. It is shown that group IVb dopants result in a slight increase in the \BM, while group IVa, except Si, and the aliovalent dopants show a decrease in the \BM\ of CeO$_2$. The decrease, however, is insufficient to obtain \BM\ matching at concentrations similar to those needed for \LP\ matching with LZO.
\section{Acknowledgement}\label{ASS:Acknowledgement}
\indent The research was financially supported by FWO-Vlaanderen, project n$^{\circ}$ $3$G$080209$, EMRS Symposium A organization and FWO (grant K1B9711N).
We acknowledge the Research Board of the Ghent University.
S.~C. acknowledges financial support from OCAS NV by an OCAS-endowed industrial chair at Ghent University. 
This work was carried out using the Stevin Supercomputer Infrastructure at Ghent University. 





\bibliographystyle{model1-num-names}






\end{document}